\documentclass[11pt,twoside]{article}
\usepackage{graphicx}
\usepackage{subfigure}
\usepackage{fancyheadings}
\usepackage{amsmath}
\usepackage{amssymb}
\usepackage{cite}
\usepackage{color,colortbl}

\definecolor{darkishgreen}{RGB}{39,203,22}
\definecolor{LightCyan}{rgb}{0.88,1,1}
\definecolor{Gray}{gray}{0.9}
\definecolor{lightRed}{RGB}{230,170,150}
\definecolor{modRed}{RGB}{230,82,90}
\definecolor{strongRed}{RGB}{230,6,6}

\begin{document}
\newcommand{\pst}{\hspace*{1.5em}}


\newcommand{\be}{\begin{equation}}
\newcommand{\ee}{\end{equation}}
\newcommand{\bm}{\boldmath}
\newcommand{\ds}{\displaystyle}
\newcommand{\bea}{\begin{eqnarray}}
\newcommand{\eea}{\end{eqnarray}}
\newcommand{\ba}{\begin{array}}
\newcommand{\ea}{\end{array}}
\newcommand{\arcsinh}{\mathop{\rm arcsinh}\nolimits}
\newcommand{\arctanh}{\mathop{\rm arctanh}\nolimits}
\newcommand{\bc}{\begin{center}}
\newcommand{\ec}{\end{center}}

\thispagestyle{plain}

\label{sh}


\begin{center} {\Large \bf
\begin{tabular}{c}
INEQUALITIES FOR PURITY PARAMETERS
\\[-1mm]
FOR MULTIPARTITE AND SINGLE QUDIT STATES
\end{tabular}
 } \end{center}

\bigskip

\bigskip

\begin{center} {\bf
V.I. Man'ko$^{1}$ and L.A. Markovich$^{2*}$
}\end{center}

\medskip

\begin{center}
{\it
$^1$P.N. Lebedev Physical Institute, Russian Academy of Sciences\\
Leninskii Prospect 53, Moscow 119991, Russia

\smallskip

$^2$Institute of Control Sciences, Russian Academy of Sciences\\
Profsoyuznaya 65, Moscow 117997, Russia\\
Moscow Institute of Physics and Technology (MIPT)
}
\smallskip

$^*$Corresponding author e-mail:~~~kimo1~@~mail.ru\\
\end{center}

\begin{abstract}\noindent
We analyze a recently found inequality for eigenvalues of the density matrix and purity parameter describing either a bipartite system state or a single qudit state.
The Minkowski type trace inequality for the density matrices of the qudit states is rewritten in terms of the purities. The properties of the obtained inequality are discussed.
  The $X$-states of the two qubits and the single qudit are considered in detail. A study of the relation of the obtained purity inequalities with the entanglement is presented.
\end{abstract}

\medskip

81P40, 81Q80
\section{Introduction}
\pst
Nowadays, the use of the density matrix operators to describe composite quantum states is of a big interest. Given the density operator $\widehat{\rho}(1,2)$ of the whole system there is a possibility to construct two reduced density operators $\widehat{\rho}(1)$ and $\widehat{\rho}(2)$ that describe states of the first and the second subsystems, respectively. Different characteristics of quantum correlations in the composite systems are well known and studied in this formalism. The properties of these correlations are reflected by the existence of the set of equalities and inequalities regarding entropies and informations, both in the quantum and classical domains \cite{Lieb,Ruskai,Winter,Petz2}.
\par However, the notion of the entanglement for the system without subsystems, i.e. the single qudit, is not so obvious. Recently,  it was observed in \cite{Chernega,Chernega:14,Manko,OlgaMankoarxiv} that the quantum properties of the systems without subsystems can be formulated by using an invertible map of integers $1,2,3\ldots$ onto the pairs (triples, etc) of integers $(i,k)$, $i,k=1,2,\ldots$ (or semiintegers). For example, the single qudit state $j=0,1/2,1,3/2,2,\ldots$ can be mapped onto the density operator of the system containing the subsystems like the state of the two qudits. Using this mapping, the notion of the separability and the entanglement was extended in \cite{Markovich3} to the case of the single qudit $X$-state with $j=3/2$.
In \cite{Cher} the Minkowski type inequality with one parameter \cite{Lieb1} for the density matrices of the bipartite system
was generalized to the case of the single qudit state which does not contain subsystems. In \cite{Markovich5} the Minkowski type trace inequality with two parameters \cite{Lieb2} was introduced for the system without subsystems. Note that the quantum correlations, similar to correlations analogous to the entanglement  for the single qudit, were found and formulated in terms of a quantum contextuality in \cite{Klichko}.
\par In \cite{Manko355} the new inequality for the purity parameters was introduced for the qudit state. The aim of the paper is to rewrite the  Minkowski type  inequality with two parameters in terms of the purities and to compare them with those given in \cite{Manko355}. The properties of the  new purity inequalities are studied on the example of the $X$-states  \cite{Hedemann}. The relation of the obtained purity inequalities with the entanglement is shown by the  example of the single qudit Gisin \cite{Gisin1}, Werner \cite{Werner} and $\beta$ states \cite{Roa}.
\section{The purity and Minkowski type trace inequality}
Let us define the $N\times N$ density matrix of the quantum state in the block form
\begin{eqnarray}\rho={\left(
                     \begin{array}{cccc}
                       a_{11} &a_{12} & \cdots & a_{1n}\\
                       a_{21} & a_{22}& \cdots & a_{2n} \\
                       \vdots & \vdots & \ddots & \vdots \\
                       a_{n1} & a_{n2} & \cdots & a_{nn} \\
                     \end{array}
                   \right)}\,, \label{3}
\end{eqnarray}
where $\rho=\rho^{\dagger}$, $Tr\rho=1$, $\rho\geq0$ and $N=nm$. The blocks $a_{ij}$ denote a $m\times m$ matrices.
If the density matrix \eqref{3} describes the system with subsystems, i.e. the  bipartite state (the two-qubit system), then we can consider two subsystems. Reduced density matrices $\rho_1$, $\rho_2$ are defined as partial traces  of \eqref{3}.
\\ However, the density matrix written in the form \eqref{3} can also describe the single qudit state. Using techniques described in  \cite{Manko355}, i.e. the analog of the partial tracing of the block matrix of the two-qubit state, we can write two matrices
\begin{eqnarray}\rho_1={Tr_2\rho=\left(
                     \begin{array}{cccc}
                       Tra_{11} &Tra_{12} & \cdots & Tra_{1n}\\
                       Tra_{21} & Tra_{22}& \cdots & Tra_{2n} \\
                       \vdots & \vdots & \ddots & \vdots \\
                       Tra_{n1} & Tra_{n2} & \cdots & Tra_{nn} \\
                     \end{array}
                   \right)},\quad \rho_2=Tr_1\rho=\sum\limits_{k=1}^{n}a_{kk}\,. \label{4}
\end{eqnarray}
The latter matrices do not correspond to subsystems like in the bipartite state, but we can formally write the purity parameter for the single qudit state by
\begin{eqnarray}\label{1}\mu_{12}=Tr \rho^2
\end{eqnarray}
and the purity parameters corresponding to the matrices \eqref{4} by
\begin{eqnarray}\label{2}\mu_{1}=Tr \rho_1^2, \quad \mu_{2}=Tr \rho_2^2.
\end{eqnarray}
In \cite{Manko355} the deformed inequality for the qudit state
\begin{eqnarray}\label{5}\mu_{1}+ \mu_{2}-1\leq\mu_{12}
\end{eqnarray}
was introduced.
For the given real numbers $p$ and $q$ we introduce the following notations
\begin{eqnarray*}&&\rho^p=\left(
                     \begin{array}{cccc}
                       a_{11}(p) &a_{12}(p) & \cdots & a_{1n}(p)\\
                       a_{21}(p) & a_{22}(p)& \cdots & a_{2n}(p) \\
                       \vdots & \vdots & \ddots & \vdots \\
                       a_{n1}(p) & a_{n2}(p) & \cdots & a_{nn}(p) \\
                     \end{array}
                   \right),\quad
                   \rho^q=\left(
                     \begin{array}{cccc}
                       a_{11}(q) &a_{12}(q) & \cdots & a_{1n}(q)\\
                       a_{21}(q) & a_{22}(q)& \cdots & a_{2n}(q) \\
                       \vdots & \vdots & \ddots & \vdots \\
                       a_{n1}(q) & a_{n2}(q) & \cdots & a_{nn}(q) \\
                     \end{array}
                   \right).
\end{eqnarray*}
The Minkowski type trace inequality with two parameters \cite{Lieb2} for the quantum system with the matrix \eqref{3} reads as
\begin{eqnarray}{\left(Tr_2\left[\left(Tr_1\rho^q\right)^{p/q}\right]\right)^{1/p}\leq\left(Tr_1\left[\left(Tr_2\rho^p\right)^{q/p}\right]\right)^{1/q}}\,. \label{18}
\end{eqnarray}
or
\begin{eqnarray}{\left(Tr\left(\sum\limits_{i=1}^na_{ii}(q)\right)^{\frac{p}{q}}\right)^{\frac{1}{p}}\leq
\left(Tr\left(
                     \begin{array}{cccc}
                       Tr a_{11}(p) & Tr a_{12}(p) & \cdots & Tr a_{1n}(p)\\
                       Tr a_{21}(p) & Tr a_{22}(p)& \cdots & Tr a_{2n}(p) \\
                       \vdots & \vdots & \ddots & \vdots \\
                       Tr a_{n1}(p) & Tr a_{n2}(p) & \cdots & Tr a_{nn}(p) \\
                     \end{array}
                   \right)^{\frac{q}{p}}\right)^{\frac{1}{q}}}\,. \label{18}
\end{eqnarray}
The inequality holds for all $1\leq q\leq p$. Otherwise, it should be reversed. The latter inequality holds both for the quantum systems with subsystems and for the single qudit states.
\par We rewrite the  inequality \eqref{18} in terms of the purities \eqref{2}. In case that $q=2$ and $p=1$, it is easy to see that the latter inequality can be rewritten as
\begin{eqnarray}\label{6}
\mu_{2}^{1/2}\leq Tr_1\left[\left(Tr_2\rho^2\right)^{1/2}\right].
\end{eqnarray}
Similarly, in the case $q=1$ and $p=2$, the inequality \eqref{18} can be rewritten as
\begin{eqnarray}\label{8}
\mu_{1}^{1/2}\leq Tr_2\left[\left(Tr_1\rho^2\right)^{1/2}\right].
\end{eqnarray}
Notice, that since $1\leq p\leq q$ holds the inequality \eqref{18} has been reversed. Combining \eqref{6} and \eqref{8} we can write
 \begin{eqnarray}\label{9}
\mu_{1}+\mu_{2}\leq \left(Tr_2\left[\left(Tr_1\rho^2\right)^{1/2}\right]\right)^2+\left(Tr_1\left[\left(Tr_2\rho^2\right)^{1/2}\right]\right)^2.
\end{eqnarray}
Reducing \eqref{9} to the form of \eqref{5}, the following inequality can be rewritten as
 \begin{eqnarray}\label{10}
\mu_{1}+\mu_{2}-1\leq \widetilde{\mu},
\end{eqnarray}
where we  introduce the notation $\widetilde{\mu}\equiv\left(Tr_2\left[\left(Tr_1\rho^2\right)^{1/2}\right]\right)^2+\left(Tr_1\left[\left(Tr_2\rho^2\right)^{1/2}\right]\right)^2-1$.
Hence, it is interesting to compare the  right hand sides of the inequalities \eqref{5} and \eqref{10}.
\section{The $X$-state for the single qudit}
If $\rho_{12}=\rho_{13}=\rho_{21}=\rho_{31}=\rho_{24}=\rho_{34}=\rho_{42}=\rho_{43}=0$ holds, then the density matrix \eqref{3} has the form of the $X$-state density matrix
                               \begin{eqnarray}\label{7}
\rho^{X}&=&\left(
                     \begin{array}{cccc}
                       \rho_{11} & 0 & 0 & \rho_{14}\\
                       0 & \rho_{22}& \rho_{23} & 0 \\
                       0 & \rho_{32} & \rho_{33} & 0 \\
                       \rho_{41} & 0 & 0 & \rho_{44} \\
                     \end{array}
                   \right)=\left(
                     \begin{array}{cccc}
                       \rho_{11} & 0 & 0 & \rho_{14}\\
                       0 & \rho_{22}& \rho_{23} & 0 \\
                       0 & \rho_{23}^{\ast} & \rho_{33} & 0 \\
                      \rho_{14}^{\ast} & 0 & 0 & \rho_{44} \\
                     \end{array}
                   \right),
\end{eqnarray}
where $\rho_{11},\rho_{22},\rho_{33},\rho_{44}$ are positive reals and $\rho_{23},\rho_{14}$ are complex quantities.
The latter matrix has the unit trace and it is nonnegative if $\rho_{22}\rho_{33}\geq|\rho_{23}|^2$, $\rho_{11}\rho_{44}\geq|\rho_{14}|^2$ hold.
In \cite{Roa} the following two inequalities for having an entanglement are indicated as
\begin{eqnarray*}&&|\rho_{23}|\leq\sqrt{\rho_{22}\rho_{33}}<|\rho_{14}|\leq\sqrt{\rho_{11}\rho_{44}},\\
&&|\rho_{14}|\leq\sqrt{\rho_{11}\rho_{44}}<|\rho_{23}|\leq\sqrt{\rho_{22}\rho_{33}}.
\end{eqnarray*}
Obviously, only one or none can be fulfilled.
\par Let us rewrite the matrix \eqref{7} in block form. The block matrices are determined by
\begin{eqnarray*}a_{11}=\left(
                         \begin{array}{cc}
                           \rho_{11} & 0 \\
                            0 & \rho_{22} \\
                         \end{array}
                       \right),\quad a_{12}=\left(
                         \begin{array}{cc}
                           0 & \rho_{14} \\
                            \rho_{23} & 0 \\
                         \end{array}
                       \right),\quad a_{21}=\left(
                         \begin{array}{cc}
                           0 & \rho_{32} \\
                            \rho_{41} & 0 \\
                         \end{array}
                       \right),\quad a_{22}=\left(
                         \begin{array}{cc}
                           \rho_{33} & 0 \\
                            0 & \rho_{44} \\
                         \end{array}
                       \right).
\end{eqnarray*}
Hence, the purity parameters \eqref{1} and \eqref{2} of the $X$-state density matrix \eqref{7} are given by
\begin{eqnarray*}\mu_{12}&=&Tr\left(
                             \begin{array}{cc}
                               a_{11}(2) & a_ {12}(2)\\
                               a_{21}(2) & a_{22}(2) \\
                             \end{array}
                           \right)=\sum\limits_{i=1}^{4}\rho_{ii}^2+2(|\rho_{14}|^2+|\rho_{23}|^2),\\
                           \mu_1&=& Tr\left(
                             \begin{array}{cc}
                               Tra_{11} & Tra_ {12}\\
                               Tra_{21} & Tra_{22} \\
                             \end{array}
                           \right)^2=\sum\limits_{i=1}^{4}\rho_{ii}^2+2(\rho_{11}\rho_{22}+\rho_{33}\rho_{44}),\\
                            \mu_2&=&Tr(a_{11}+a_{22})^2=\sum\limits_{i=1}^{4}\rho_{ii}^2+2(\rho_{11}\rho_{33}+\rho_{22}\rho_{44}).
\end{eqnarray*}
Hence, substituting the latter purities in \eqref{5} we can write
\begin{eqnarray}\label{11}
&&2\sum\limits_{i=1}^{4}\rho_{ii}^2+2(\rho_{11}+\rho_{44})(\rho_{22}+\rho_{33})-1\leq\sum\limits_{i=1}^{4}\rho_{ii}^2+2(|\rho_{14}|^2+|\rho_{23}|^2).
\end{eqnarray}
Analogously, the inequality  \eqref{10} has the following representation
\begin{eqnarray}\label{12}
&&2\sum\limits_{i=1}^{4}\rho_{ii}^2+2(\rho_{11}+\rho_{44})(\rho_{22}+\rho_{33})-1\leq\widetilde{\mu},
\end{eqnarray}
where
\begin{eqnarray*}
\widetilde{\mu}&=&2\sqrt{\rho_{11}^2+\rho_{22}^2+|\rho_{14}|^2+|\rho_{23}|^2}\sqrt{\rho_{33}^2+\rho_{44}^2+|\rho_{14}|^2+|\rho_{23}|^2}+2\sum\limits_{i=1}^{4}\rho_{ii}^2\nonumber\\
&+&2\sqrt{\rho_{11}^2+\rho_{33}^2+|\rho_{14}|^2+|\rho_{23}|^2}\sqrt{\rho_{22}^2+\rho_{44}^2+|\rho_{14}|^2+|\rho_{23}|^2}
+4(|\rho_{14}|^2+|\rho_{23}|^2)-1.
\end{eqnarray*}
Let us introduce examples of how the parameters of the density matrices impact on the sign of $\widetilde{\mu}-\mu_{12}$ and its connection with the entanglement.
\par
\begin{figure}[ht]
\begin{center}
\begin{minipage}[ht]{0.49\linewidth}
\includegraphics[width=1\linewidth]{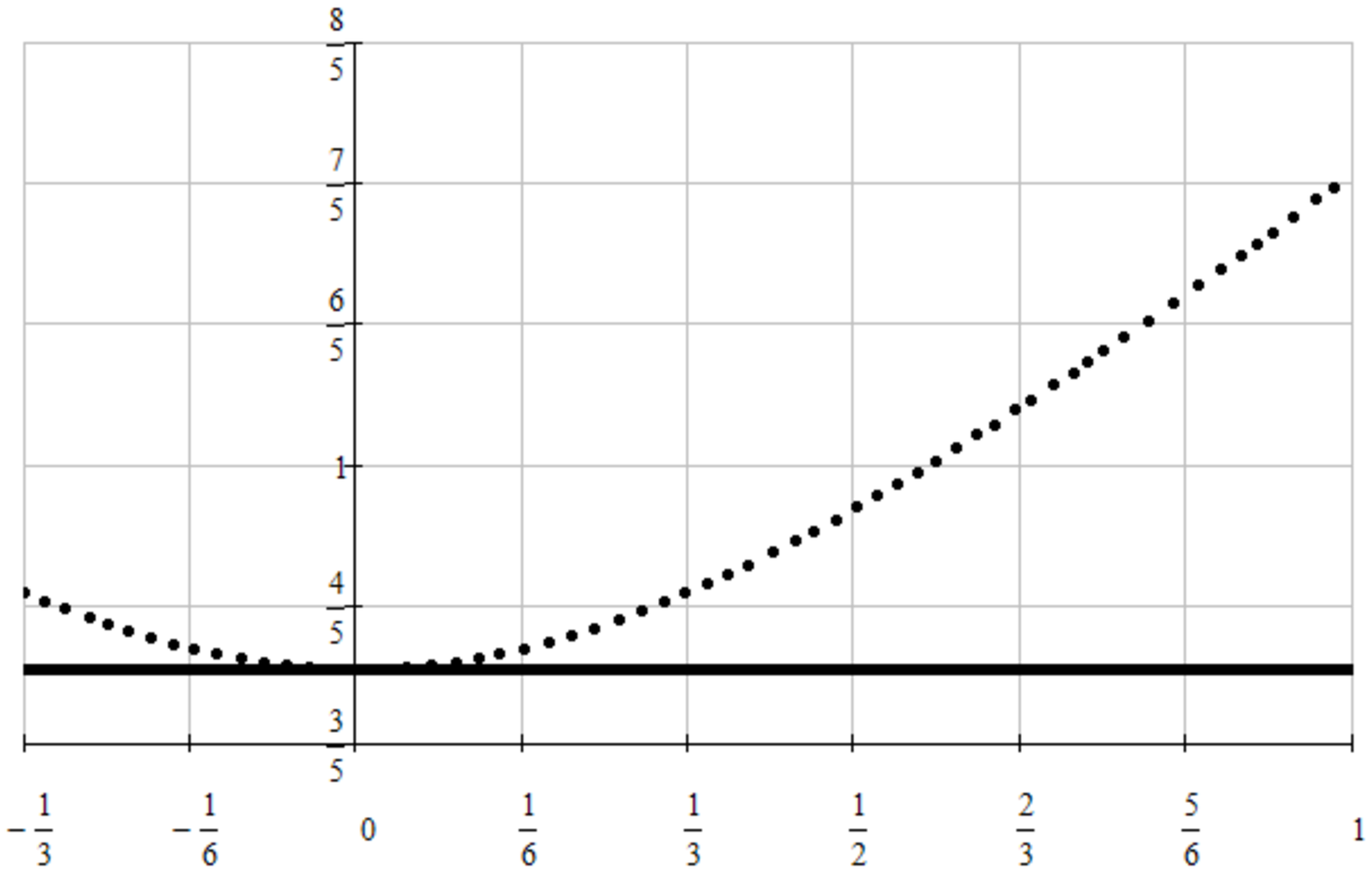}
\vspace{-4mm}
\caption{The left part (black line) and the right part (dotted line) of the inequality \eqref{6} for the Werner state  \eqref{10_1} of the qudit with the spin $j=3/2$   against the parameter $p$.}
\label{fig:1}
\end{minipage}
\hfill
\begin{minipage}[ht]{0.49\linewidth}
\includegraphics[width=1\linewidth]{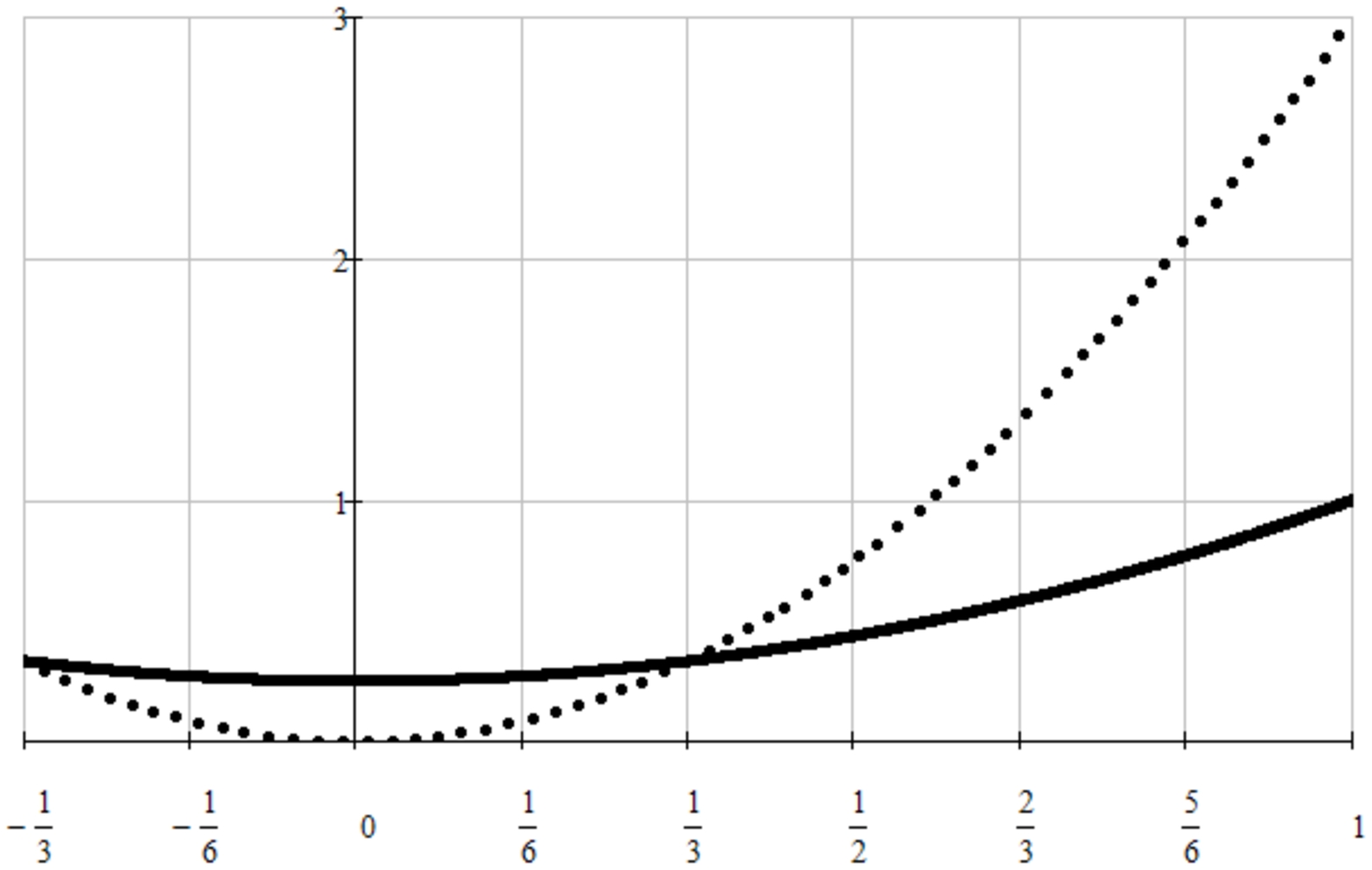}
\vspace{-4mm}
\caption{$\mu_{12}$ (black line), $\widetilde{\mu}$ (dotted line) for the Werner state  \eqref{10} of the qudit with the spin $j=3/2$  against the parameter $p$.}
\label{fig:2}
\end{minipage}
\end{center}
\end{figure}

 One of the examples of the $X$-state is given by the Werner state \cite{Werner}. The single qudit with the spin $j=3/2$ can be described by the following Werner density matrix
\begin{eqnarray}\label{10_1}\rho^{W}&=&\left(
                     \begin{array}{cccc}
                       \frac{1+p}{4} & 0 & 0 & \frac{p}{2}\\
                       0 & \frac{1-p}{4}& 0 & 0 \\
                       0 & 0& \frac{1-p}{4} & 0 \\
                       \frac{p}{2} & 0 & 0 & \frac{1+p}{4} \\
                     \end{array}
                   \right),
\end{eqnarray}
where the parameter $p$ satisfies the inequality $-\frac{1}{3}\leq p\leq1$. The parameter domain $\frac{1}{3}< p\leq1$ corresponds to the entangled state.
The purities \eqref{1} and \eqref{2} of the matrix \eqref{10_1} are determined by
\begin{eqnarray*}\mu_{12}=\frac{3p^2+1}{4},\quad  \mu_{1}=\frac{1}{2},\quad \mu_{2}=\frac{1}{2}.
\end{eqnarray*}
Hence, the inequality \eqref{5} has the form $0\leq\frac{3p^2+1}{4}$. For the Werner state the inequalities \eqref{6} and \eqref{8} are the same and state as
$\frac{1}{2}\leq \frac{\sqrt{6p^2+2}}{2}$. Hence, the inequality \eqref{10} implies $0\leq 3p^2$. The latter results are illustrated by Figures~\ref{fig:1} and \ref{fig:2} for  various values of the parameter $p$ of the matrix \eqref{10_1}.
Visible, that the right part of the inequality \eqref{10} is larger than the right part of the inequality \eqref{5}, i.e. $\widetilde{\mu}-\mu_{12}>0$, when the parameter $p$ belongs to the domain $\frac{1}{3}< p\leq1$ which corresponds to the entangled state.
\par The second example of the $X$-state is provided by the Gisin state \cite{Gisin1}. It is given by the following density matrix
\begin{eqnarray}\label{10_2}\rho^{G}&=&\left(
                     \begin{array}{cccc}
                       \frac{1-x}{2} & 0 & 0 & 0\\
                       0 & x|a|^2& xab^{\ast} & 0 \\
                       0 & xa^{\ast}b& x|b|^2 & 0 \\
                       0 & 0 & 0 & \frac{1-x}{2} \\
                     \end{array}
                   \right),
\end{eqnarray}
where $|a|^2+|b|^2=1$, $0<x<1$. In \cite{Peres} it is shown, that the latter matrix has positive eigenvalues if $x\leq\frac{1}{1+2|ab|}$ holds. We denote it as $x_{max}=\frac{1}{1+2|ab|}$.
The left hand side of \eqref{5} and \eqref{10} is equal to
\begin{eqnarray}\label{20}\mu_{1}+ \mu_{2}-1&=&x^2\left(2\left(|a|^4+|b|^4\right)-1\right).
\end{eqnarray}
The right hand side of the inequality \eqref{10} reads
 \begin{eqnarray}\label{21}
\widetilde{\mu}&=&x(3x-2)+\sqrt{x^2(4|a|^2+1)-2x+1}\sqrt{x^2(4|b|^2+1)-2x+1}.
\end{eqnarray}
The purity of the whole system is given by the following term
\begin{eqnarray}\label{22}\mu_{12}&=&\frac{3}{2}x^2-x+\frac{1}{2}.
\end{eqnarray}
 We can rewrite the inequalities \eqref{5} and \eqref{10} using the latter results. The results are shown in Figures~\ref{fig:3} and \ref{fig:4}.
\begin{figure}[ht]
\begin{center}
\begin{minipage}[ht]{0.49\linewidth}
\includegraphics[width=1\linewidth]{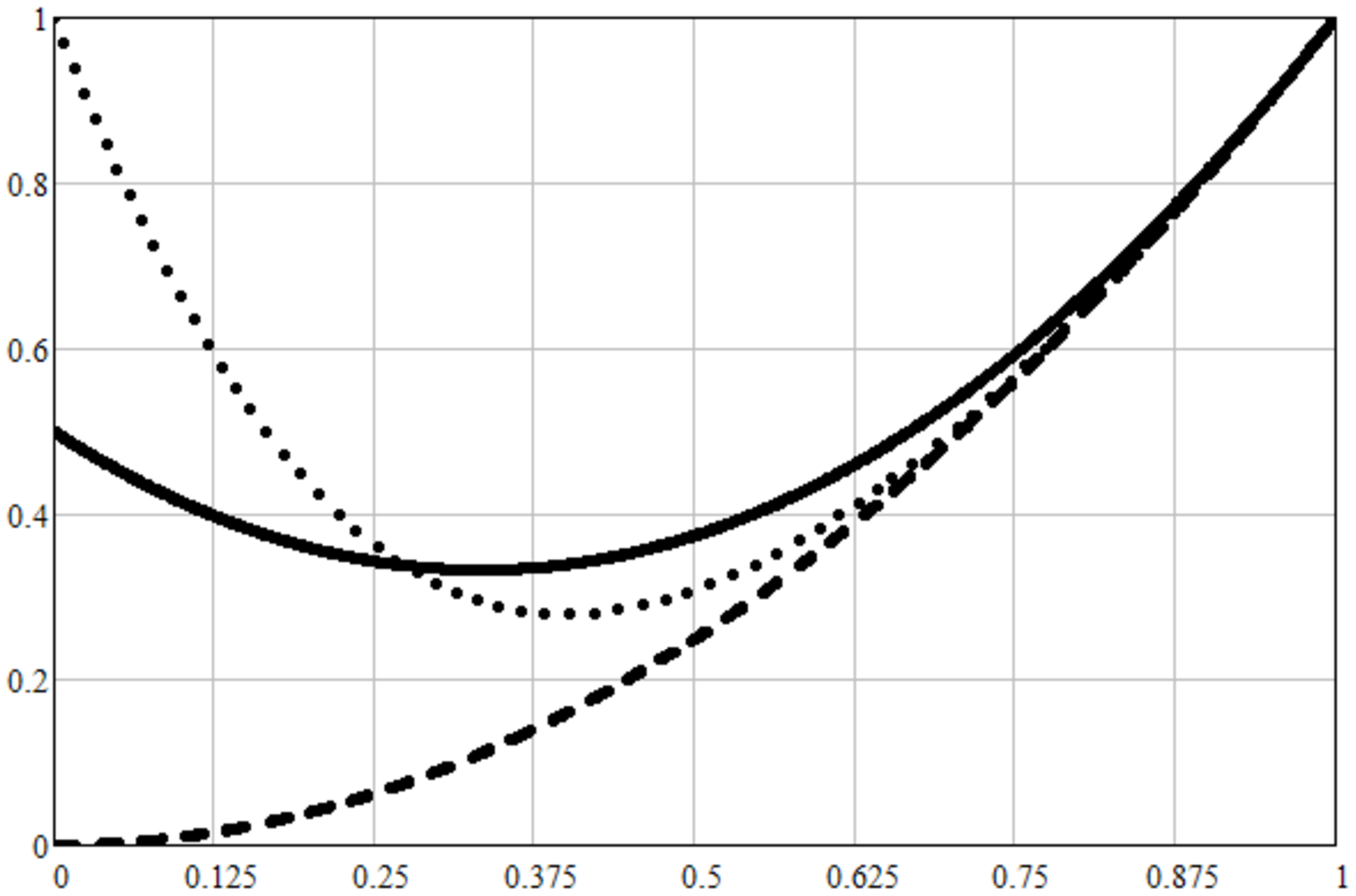}
\vspace{-4mm}
\caption{$\mu_{12}$ (black line), $\widetilde{\mu}$ (dotted line) and \eqref{20} (dashed line) for the Gisin state  \eqref{10_2} of the qudit with the spin $j=3/2$   against the parameter $x$ for $\{a,b\}=\{1,0\}$, $x_{max}=1$.}
\label{fig:3}
\end{minipage}
\hfill
\begin{minipage}[ht]{0.49\linewidth}
\includegraphics[width=1\linewidth]{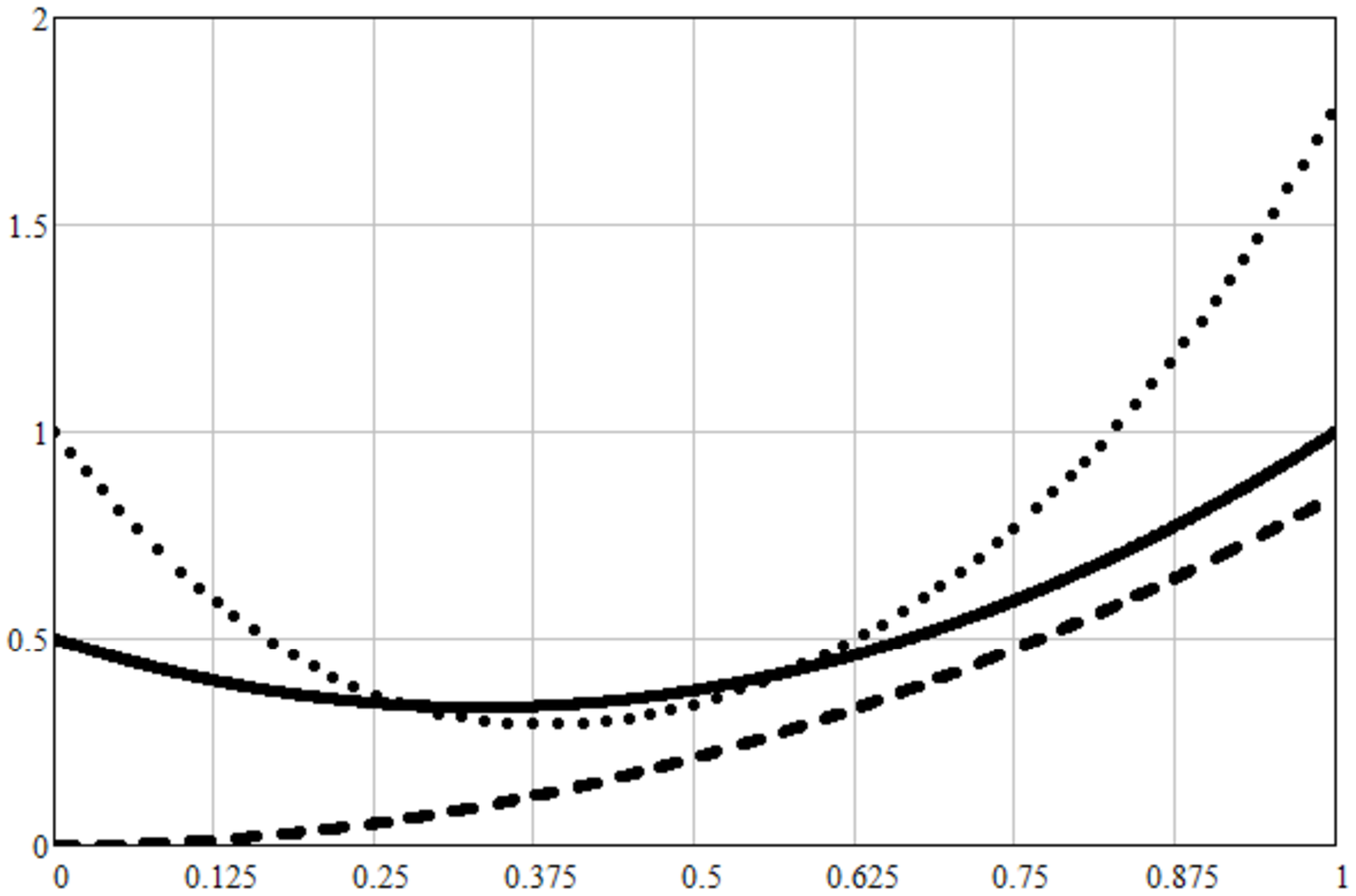}
\vspace{-4mm}
\caption{$\mu_{12}$ (black line), $\widetilde{\mu}$ (dotted line) and \eqref{20} (dashed line) for the Gisin state  \eqref{10_2} of the qudit with the spin $j=3/2$   against the parameter $x$ for $\{a,b\}\cong\{0.2,0.979\}$, $x_{max}=0.718$.}
\label{fig:4}
\end{minipage}
\end{center}
\end{figure}
\begin{figure}[ht]
\begin{center}
\begin{minipage}[ht]{0.49\linewidth}
\includegraphics[width=1\linewidth]{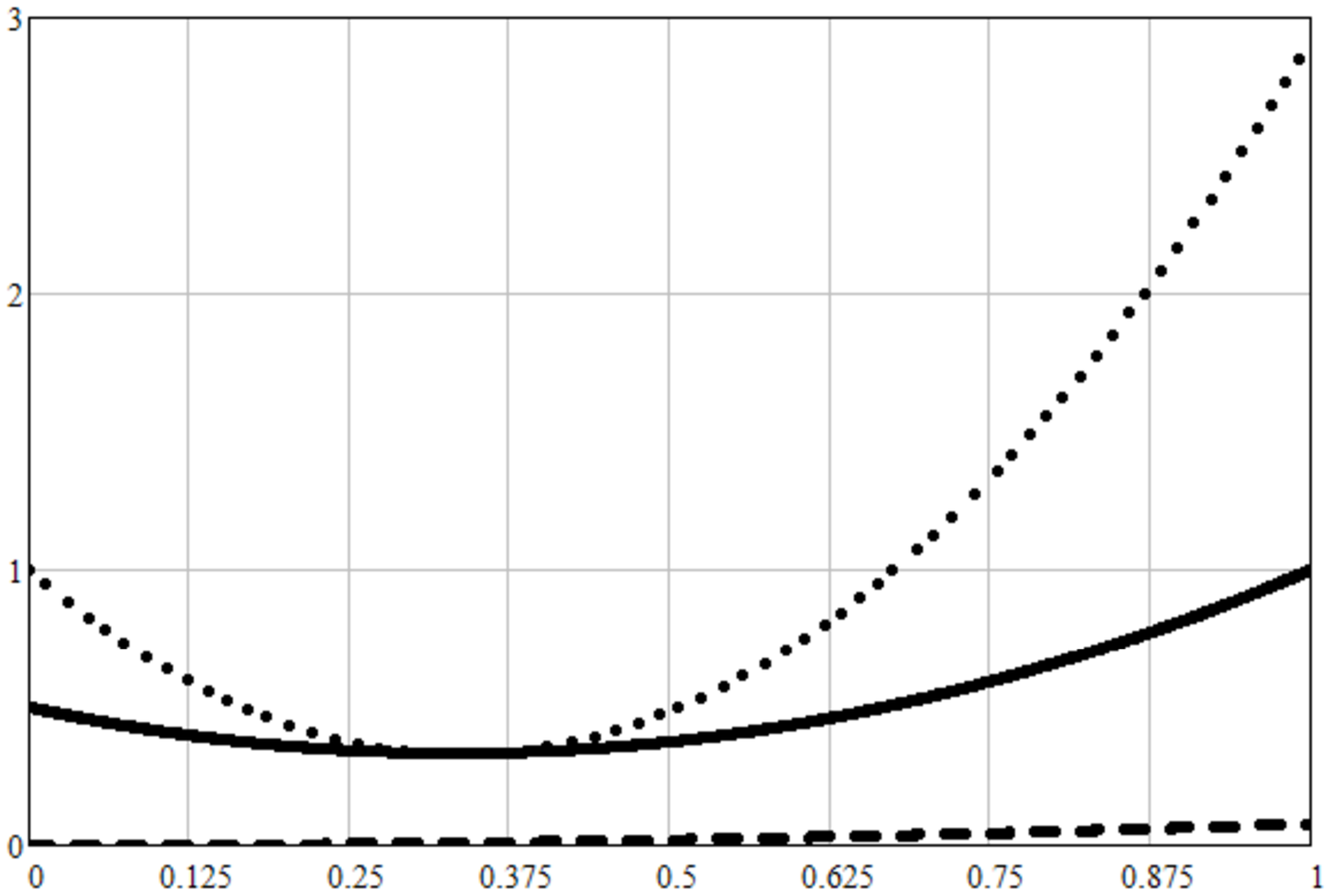}
\vspace{-4mm}
\caption{$\mu_{12}$ (black line), $\widetilde{\mu}$ (dotted line) and \eqref{20} (dashed line) for the Gisin state  \eqref{10_2} of the qudit with the spin $j=3/2$   against the parameter $x$ for $\{a,b\}=\{0.6,0.8\}$, $x_{max}=0.51$.}
\label{fig:3}
\end{minipage}
\hfill
\begin{minipage}[ht]{0.49\linewidth}
\includegraphics[width=1\linewidth]{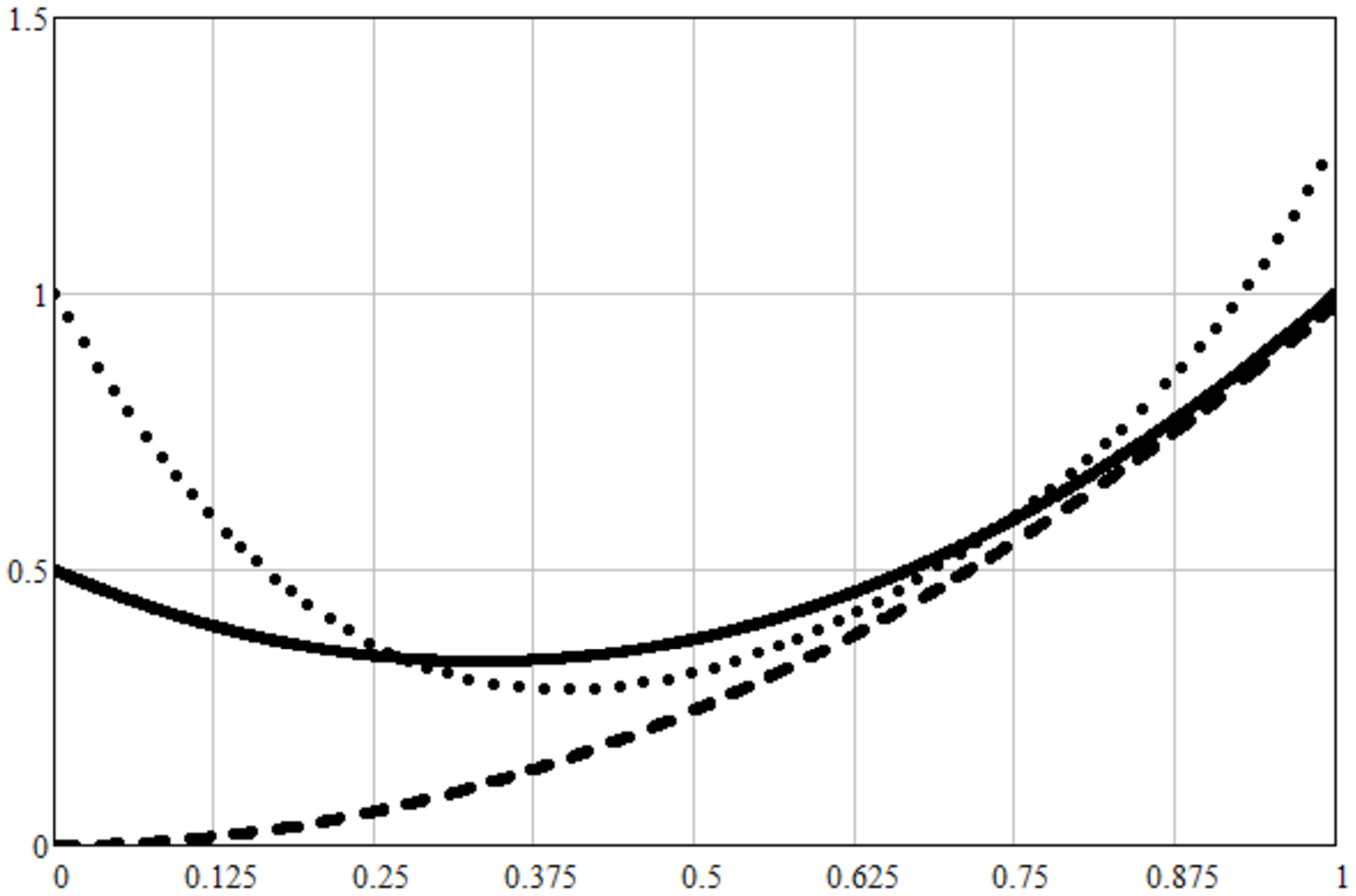}
\vspace{-4mm}
\caption{$\mu_{12}$ (black line), $\widetilde{\mu}$ (dotted line) and \eqref{20} (dashed line) for the Gisin state  \eqref{10_2} of the qudit with the spin $j=3/2$   against the parameter $x$ for $\{a,b\}\cong\{0.07,0.99\}$, $x_{max}=0.87$.}
\label{fig:4}
\end{minipage}
\end{center}
\end{figure}
The intersection points are determined by roots of the equation $\widetilde{\mu}-\mu_{12}=0$.
Similarly to the Werner state, we have $\widetilde{\mu}-\mu_{12}>0 $ for the Gisin state  in the domain $x>x_{max}$ that corresponds to the entangled state.
\par The third example of the $X$-state is provided by the $\beta$-state of the single qudit with the spin $j=3/2$. It is given by the following density matrix
\begin{eqnarray}\label{10_3}\rho^{\beta}&=&\left(
                     \begin{array}{cccc}
                       \frac{\beta}{2} & 0 & 0 & \frac{\beta}{2}\\
                       0 & \frac{1-\beta}{2}& \frac{1-\beta}{2} & 0 \\
                       0 & \frac{1-\beta}{2}& \frac{1-\beta}{2} & 0 \\
                       \frac{\beta}{2} & 0 & 0 & \frac{\beta}{2} \\
                     \end{array}
                   \right),
\end{eqnarray}
where the parameter is $0\leq\beta\leq1$. The latter state is separable in $\beta=1/2$. The purities for the $\beta$-state are given by
\begin{eqnarray*}\mu_{12}=2\beta^2-2\beta+1,\quad  \mu_{1}=\frac{1}{2},\quad \mu_{2}=\frac{1}{2}
\end{eqnarray*}
and the right hand side of the inequality \eqref{10} reads
 \begin{eqnarray*}\widetilde{\mu}=8\beta^2-8\beta+3.
\end{eqnarray*}
\begin{figure}[ht]
\begin{center}
\begin{minipage}[ht]{0.49\linewidth}
\includegraphics[width=1\linewidth]{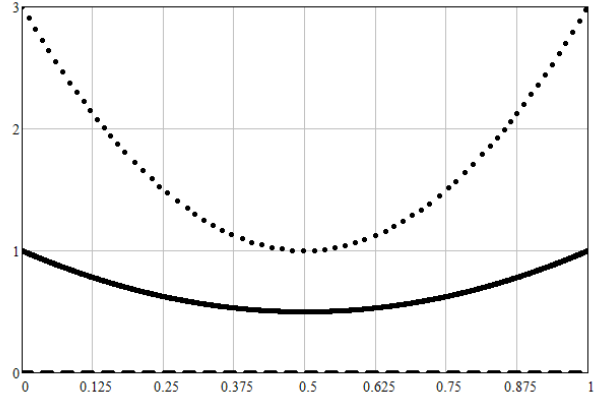}
\vspace{-4mm}
\caption{The purity $\mu_{12}$ (black line), $\widetilde{\mu}$ (dotted line) and \eqref{20} (dashed line) for the $\beta$-state  \eqref{10_3} of the qudit with the spin $j=3/2$   against the parameter $\beta$.}
\label{fig:5}
\end{minipage}
\end{center}
\end{figure}
The comparison of $\mu_{12}$ and $\widetilde{\mu}$ is shown in Figure~\ref{fig:5}. Since the state is entangled in the whole domain except of the point $\beta=1/2$, the
difference between the right hand sides of the inequalities \eqref{5} and \eqref{10} is positive for all $\beta$, i.e.  $\widetilde{\mu}-\mu_{12}>0$.

 \section{Summary}
\pst
To conclude we point out the main results of the work. The Minkowski type trace inequality was rewritten in terms of the purities. Two new inequalities were obtained and compared with the known inequality for the purities given in \cite{Manko355}. It was shown by the examples of the Werner, the Gisin and the $\beta$ states that for the entangled state the right hand side of the inequality \eqref{10} is higher than the right hand side of the inequality \eqref{5}. The obtained results give an idea, that in those domains where the quantum state is entangled the difference $\widetilde{\mu}-\mu_{12}$ is necessarily positive.

\end{document}